\documentclass[
]{ceurart}

\usepackage{listings}
\usepackage{graphicx}
\usepackage{textcomp}
\usepackage{xcolor}
\usepackage{multirow} 
\usepackage{booktabs} 
\usepackage{subcaption} 
\usepackage{graphicx}

\usepackage{hyperref}

\usepackage{url}
\begin{document}

\copyrightyear{2025}
\copyrightclause{Copyright for this paper by its authors.
  Use permitted under Creative Commons License Attribution 4.0
  International (CC BY 4.0).}

\conference{}
\title{Embedding-Aware Quantum-Classical SVMs for Scalable Quantum Machine Learning}

\author[1]{Sebastián Andrés Cajas Ordóñez}[%
orcid=0000-0003-0579-6178,
email=sebastian.cajasordonez@ucd.ie,
]
\cormark[1]
\address[1]{National Irish Centre for AI (CeADAR), University College Dublin (UCD), Dublin, Ireland}

\author[2]{Luis Fernando Torres Torres}[%
orcid=0009-0001-0706-0626,
email=lf.torres@udea.edu.co,
]
\address[2]{SISTEMIC Research Group, University of Antioquia, Medellín, Colombia}

\author[3]{Mario Bifulco}[%
email=mario.bifulco@unito.it,
]
\address[3]{Department of Computer Science, University of Torino, Torino, Italy}

\author[4]{Carlos Andres Duran}[%
orcid=0009-0008-3243-7684,
email=carlos.duran@unicauca.edu.co,
]
\address[4]{Corporation for Aerospace Initiatives (CASIRI), University of Cauca, Popayán, Colombia}

\author[1]{Cristian Bosch}[%
email=cristian.boschserrano@ucd.ie,
]

\author[1]{Ricardo Simon Carbajo}[%
orcid=0000-0002-2121-2841,
email=ricardo.simoncarbajo@ucd.ie,
]

\cortext[1]{Corresponding author.}
\begin{abstract}
Quantum Support Vector Machines face scalability challenges due to high-dimensional quantum states and hardware limitations. We propose an embedding-aware quantum-classical pipeline combining class-balanced \textit{k-means} distillation with pretrained Vision Transformer embeddings. Our key finding: ViT embeddings uniquely enable quantum advantage, achieving up to 8.02\% accuracy improvements over classical SVMs on Fashion-MNIST and 4.42\% on MNIST, while CNN features show performance degradation. Using 16-qubit tensor network simulation via cuTensorNet, we provide the first systematic evidence that quantum kernel advantage depends critically on embedding choice, revealing fundamental synergy between transformer attention and quantum feature spaces. This provides a practical pathway for scalable quantum machine learning that leverages modern neural architectures.
\end{abstract}


\maketitle

\section{Introduction}
Quantum computing has emerged as a transformative paradigm with the potential to outperform classical approaches on specialized computational tasks \cite{bausch2024learning}. Concurrently, machine learning (ML) continues advancing rapidly, driven by increasing data availability and accelerated computing hardware \cite{peral2024systematic}. Quantum machine learning (QML), at the intersection of these fields, has significant potential to unlock new capabilities in data processing and complex problem solving \cite{peral2024systematic,havlivcek2019supervised}. By utilizing quantum phenomena such as superposition and entanglement, QML algorithms may effectively address high dimensionality and combinatorial complexity beyond classical counterparts \cite{abbas2021power,senokosov2024quantum}.

Despite these prospects, large-scale quantum computing faces substantial practical challenges from noise and decoherence. Advanced error correction protocols can address these issues through sophisticated decoding algorithms, though their design complexity has led researchers to increasingly employ machine learning approaches for automation and enhancement of these decoding tasks \cite{bausch2024learning}. Beyond error correction, QML applications have been explored extensively across domains such as image classification, natural language processing, and high energy physics \cite{shen2024classification,don2024fusion,belis2024guided,vasques2023application,kim2023classical}. Although proof of concept demonstrations on quantum processors like Google's Sycamore \cite{huang2022quantum} and IBM's superconducting systems \cite{gentinetta2024complexity,cugini2023comparing} show feasibility, significant gaps persist between laboratory experiments and reliable industrial deployment \cite{gujju2024quantum,basilewitsch2024quantum}.

A promising strategy for overcoming these gaps involves leveraging embedding and dimensionality reduction methods. Classical preprocessing techniques such as principal component analysis, or learned neural encoders like variational autoencoders, effectively reduce dataset complexity prior to quantum model input \cite{belis2024guided,phalak2024optimizing}. These hybrid approaches help manage limited qubit resources on current quantum hardware while exploiting quantum enhanced feature spaces \cite{beer2020training,senokosov2024quantum}. Benchmark studies comparing quantum and classical ML approaches underline the potential and present limitations of QML \cite{havenstein2018comparisons,gujju2024quantum,basilewitsch2024quantum}. Advancing this field requires systematic evaluations under realistic conditions, incorporating representative datasets, accurate noise models, and relevant performance metrics \cite{peral2024systematic,havlivcek2019supervised,potempa2022comparing}.

We propose an embedding-aware, hybrid quantum-classical QSVM framework designed to address the scalability limitations of quantum machine learning. By integrating class-balanced $k$-means data distillation with pretrained embeddings, our pipeline reduces data dimensionality while preserving task-relevant structure. Quantum kernel classification is performed using tensor network simulation with NVIDIA’s cuTensorNet~\cite{cutensornet, chen2024validating}. Benchmarking on MNIST and Fashion-MNIST shows that this embedding-driven approach consistently outperforms classical and quantum baselines in both accuracy and efficiency, confirming its value for scalable and resource-constrained quantum machine learning applications.

\section{Related Work}

\subsection{Scaling QML with Simulation Frameworks}

Simulating larger circuits remains a popular strategy because near-term quantum devices have limited qubits. Efficient tensor-network methods can push simulations of quantum support vector machines (QSVMs) to hundreds of qubits \cite{chen2024validating}, addressing scaling issues that plague naïve state-vector simulators. This line of research proves instrumental for prototyping advanced QML algorithms, guiding their eventual deployment on real hardware \cite{phalak2024optimizing,mitarai2018quantum}. %

\subsection{Quantum Classifiers for Image Recognition}

Variational quantum classifiers (VQCs), variational quantum circuits~\cite{shen2024classification,farhi2018classification,mitarai2018quantum}, quantum kernels, and hybrid architectures~\cite{don2024fusion,kim2023classical,havlivcek2019supervised,sharma2023role,chen2022generating,blank2020quantum} have been applied to standard benchmarks such as MNIST, Fashion-MNIST, and medical imaging tasks~\cite{shen2024classification,don2024fusion,vasques2023application,maouaki2024quantum}. Although classical deep networks often outperform small quantum models on large-scale datasets, quantum classifiers show competitive performance in data-scarce or high-dimensional settings by leveraging specialized embeddings and kernel methods~\cite{belis2024guided,blank2020quantum}. Several studies have also demonstrated end-to-end quantum classification pipelines executed on actual hardware, though typically limited to smaller datasets due to current device constraints~\cite{shen2024classification,kim2023classical}.

\subsection{Hybrid Classical-Quantum Techniques}

Several researchers propose hybrid approaches: classical layers for data preprocessing or encoding, followed by quantum layers for feature transformation or classification \cite{don2024fusion,belis2024guided,phalak2024optimizing}. These methods can offset limited qubit counts by handing only compressed or task-relevant information to the quantum circuit \cite{shen2024classification}. Genetic algorithms \cite{phalak2024optimizing}, autoencoders \cite{belis2024guided}, and transfer learning \cite{kim2023classical} have all been employed to optimize these hybrid models.

\subsection{Applications Beyond Image Classification}

QML has also been trialed in domains like high-energy physics \cite{belis2024guided,cugini2023comparing}, medical diagnosis \cite{don2024fusion,maouaki2024quantum}, and scientific computing \cite{chen2024validating}. In certain specialized tasks - e.g., identifying the Higgs boson in proton collision data \cite{belis2024guided} - quantum models can match or exceed classical baselines under realistic noise models. These studies highlight the versatility of QML but also the pressing need for systematic benchmarking to compare cost-benefit trade-offs \cite{havenstein2018comparisons,basilewitsch2024quantum}.

As a whole, prior works illustrate QML's broad applicability, from error decoding \cite{bausch2024learning} and fundamental reviews \cite{peral2024systematic,gujju2024quantum} to specialized classifiers for real-world tasks \cite{shen2024classification,don2024fusion,kim2023classical}. However, two critical gaps remain in the literature. First, a cohesive comparison that unifies insights across multiple domains and positions these results against robust classical baselines remains a key frontier \cite{gujju2024quantum,basilewitsch2024quantum}. Second, existing approaches lack systematic investigation of how different embedding strategies affect quantum advantage, particularly the synergy between modern neural representations and quantum feature spaces. 

The present study addresses both gaps by systematically evaluating quantum-enhanced classification models alongside classical baselines using diverse embedding strategies, revealing fundamental relationships between representation choice and quantum kernel performance on representative datasets.

\section{Methodology}

\subsection{Strategy Overview}

\begin{figure*}[ht!]
  \centering
  \includegraphics[width=0.8\textwidth]{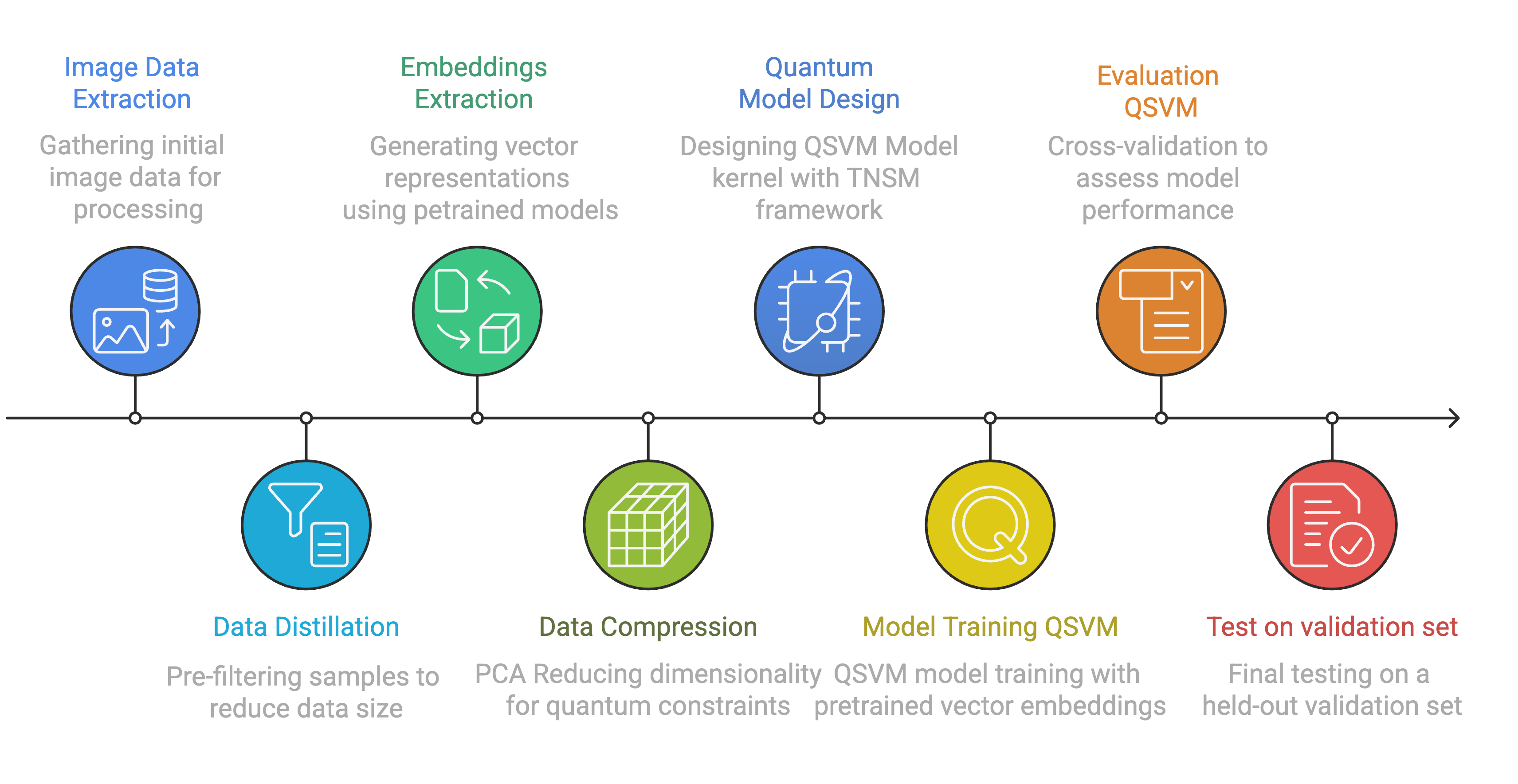}
  \caption{Illustrates the sequential steps from data extraction to QSVM evaluation. The process begins with image data extraction, followed by class-balanced $k$-means clustering to distill representative samples. Vector embeddings are then extracted using ImageNet-pretrained models such as EfficientNet or ViT. To reduce dimensionality and match quantum hardware constraints, PCA is applied to compress the embeddings. These processed embeddings are used to design a Quantum Support Vector Machine (QSVM) using the TNSM framework, which constructs a quantum kernel via a data re-uploading and compute–uncompute strategy. The model is trained and validated through cross-validation, then evaluated on a held-out test set.}
  \label{fig:pipeline}
\end{figure*}

Our approach addresses the scalability challenges of quantum machine learning through a hybrid quantum-classical pipeline that strategically combines data preprocessing, feature extraction, and quantum kernel methods. As illustrated in Figure~\ref{fig:pipeline}, the framework operates through eight sequential stages: we begin with image data extraction and preprocessing, followed by data distillation using class-balanced $k$-means clustering to reduce dataset size while maintaining representative samples. Next, we generate vector representations using pretrained models such as EfficientNet-B3~\cite{tan2019efficientnet} and Vision Transformer variants~\cite{dosovitskiy2020image}, then apply Principal Component Analysis (PCA) to compress embeddings and match quantum hardware constraints. The processed embeddings are used to design a Quantum Support Vector Machine (QSVM) using the Tensor Network Support Matrix (TNSM) framework~\cite{chen2024validating}, which constructs quantum kernels through parameterized circuits and tensor network simulation using a data re-uploading and compute-uncompute strategy. Finally, we evaluate model performance through cross-validation and test on a held-out validation set to assess generalization capability. This embedding-aware strategy enables us to leverage the representational power of modern neural architectures while exploiting quantum kernel advantages for classification tasks, making quantum machine learning more practical and scalable for real-world applications.

\subsection{Quantum Model Architecture}
\label{sec:qs_backend}
\begin{figure}[ht!]
  \centering
  \begin{subfigure}[b]{\linewidth}
    \centering
    \includegraphics[width=0.8\linewidth]{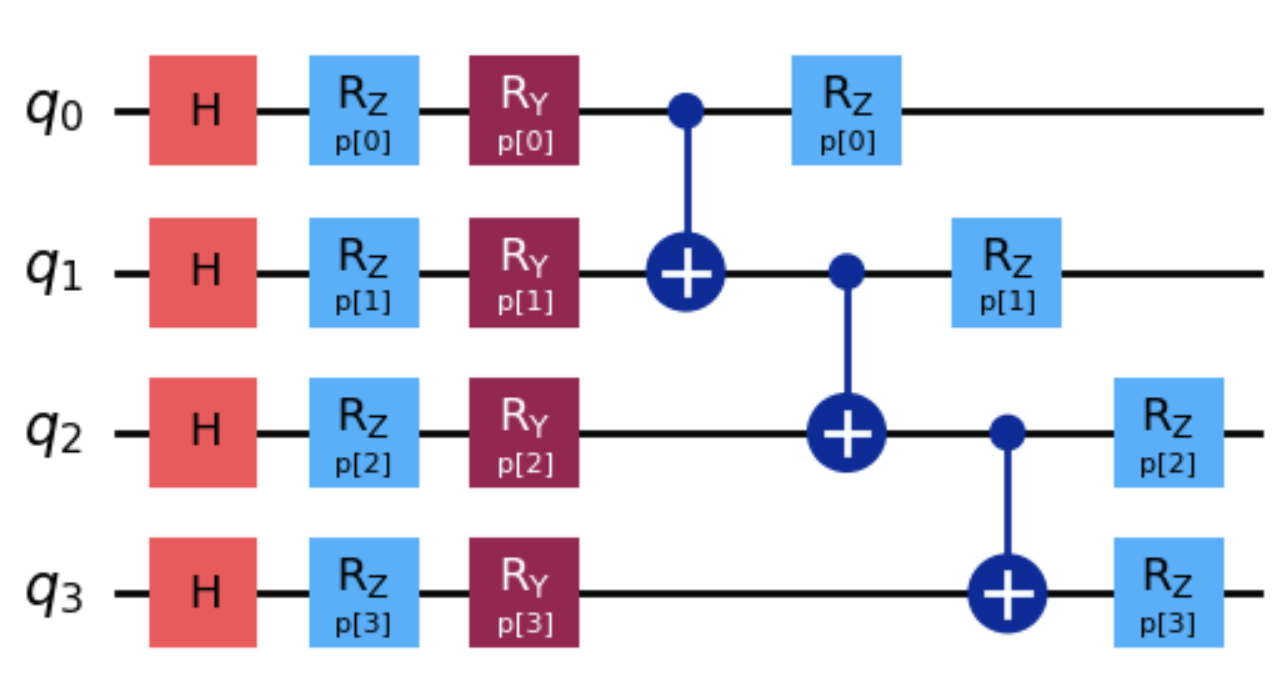} 
  \end{subfigure}
  \caption{Quantum circuit used in the QSVM pipeline using Qiskit. Each of the four qubits is initialized with a Hadamard gate, followed by data-encoding rotations using parameterized $R_Z$ and $R_Y$ gates. A sequence of CNOT gates creates entanglement between adjacent qubits, after which a second layer of $R_Z$ gates is applied. This structure forms an embedding-aware quantum feature map for encoding classical input features.\label{fig:circuit}}
\end{figure}

Chen et al.~\cite{chen2024validating} forms the foundational simulation framework upon which our quantum model is built. The architecture employs a parameterized quantum circuit that encodes input data through rotational gates and entanglement layers within a Block-Encoded State (BPS) circuit, as shown in Figure~\ref{fig:circuit}. This data re-uploading circuit design has been validated in quantum learning applications and implemented within the Qiskit framework~\cite{javadi2024quantum}, serving as a proven foundation for kernel-based quantum classification.

It constructs a quantum kernel using a compute–uncompute strategy applied to the BPS circuit, parameterized on the input examples.
This method enables the model to capture complex relationships between data points effectively.
The resulting circuits are mapped onto tensor networks using the CircuitToEinsum converter, enabling efficient simulation on classical hardware.
The kernel matrix is computed by contracting the tensor networks for all training pairs using an autotuned contraction path.

\textbf{Implementation Optimizations.}
In contrast to the original implementation~\cite{chen2024validating}, we introduce several performance enhancements to the operand construction pipeline.
First, to reduce redundant computations of trigonometric and exponential operations across repeated input angles, we apply function-level caching using Python's \texttt{@cache} decorator.
This memorization strategy significantly reduces overhead during the generation of gate matrices (e.g., parameterized $Z$ and $Y$ gates), which are frequently reused across multiple operand evaluations.
Additionally, we precompute sine and cosine values in a shared utility function to avoid duplicating expressions and improve code modularity.
Operand batches are generated via list comprehensions instead of iterative \texttt{append} calls, enhancing memory efficiency and readability.
Likewise, tensor network amplitudes are computed using preallocated lists, eliminating dynamic resizing and reducing garbage collection pressure.

These optimizations form our enhanced baseline, designated as \textit{Baseline+}, while the original implementation is referred to as \textit{ Baseline}. All comparisons throughout this paper use Baseline+ as the reference. The improvements collectively reduce execution time and peak memory usage during simulation, as demonstrated in Table~\ref{tab:results_mnist_fmnist_full}, and establish a foundation for future CuPy-based parallelization in operand template generation.


\subsection{Quantum Kernel Formulation}

The quantum kernel between two data points $x_i$ and $x_j$ is computed as the transition amplitude between their corresponding quantum states \cite{havlivcek2019supervised, schuld2019quantum}:

\begin{equation}
K_q(x_i, x_j) = |\langle \phi(x_i) | \phi(x_j) \rangle|^2
\end{equation}

where $|\phi(x)\rangle = U(x)|0\rangle^{\otimes n}$ represents the quantum feature map implemented by our parameterized circuit $U(x)$. 
The quantum advantage emerges from the exponentially large Hilbert space dimension $2^n$ compared to the classical feature space \cite{abbas2021power}. However, this advantage is critically dependent on how classical data $x$ is embedded before quantum encoding.

\section{Experimental Setup}

\subsection{Dataset and Data Distillation Process}

Due to the computational complexity of scaling Quantum Support Vector Machines (QSVMs) to high-dimensional data, particularly in the context of image classification, we adopt a distilled version of the dataset to reduce resource requirements while preserving performance.

\textbf{MNIST Dataset:} We use the MNIST dataset \cite{lecun1998gradient}, a widely recognized benchmark for evaluating image classification models. It consists of 70,000 grayscale images of handwritten digits (0–9), each of size $28 \times 28$ pixels, divided into 60,000 training and 10,000 test samples.

\textbf{Fashion-MNIST Dataset:} We also utilize the Fashion-MNIST dataset \citep{xiao2017fashion}, a benchmark dataset designed as a more challenging alternative to MNIST. It comprises 70,000 grayscale images of 10 fashion item categories (e.g., t-shirts, trousers, dresses), each of size $28 \times 28$ pixels, split into 60,000 training and 10,000 test samples. This dataset provides a diverse set of visual patterns, enabling robust evaluation of our QSVM pipeline in a multi-class classification setting.

\textbf{Data Distillation:} To address QSVM scalability constraints, we employ a class-balanced dataset distillation approach based on $k$-means clustering. The algorithm iterates through each class, applies $k$-means with $k=200$ to identify representative centroids, and selects the real data point closest to each centroid as a prototype, yielding exactly 200 samples per class. The resulting distilled dataset contains 2,000 samples total (1,600 for training, 400 for testing), reducing computational complexity from $\mathcal{O}(70000^2)$ to $\mathcal{O}(1600^2)$ kernel evaluations while preserving representative coverage of each class's feature distribution and eliminating class imbalance effects. The distillation parameters ($k$ value and dataset size) can be customized in our implementation based on available computational resources and hardware constraints, enabling adaptation to different quantum simulation capabilities.

\subsection{Embedding Extraction and Data Compression}

To construct compact and informative inputs for quantum classification, we extract high-dimensional feature embeddings using pretrained image encoders. Specifically, we employ EfficientNet-B3~\cite{tan2019efficientnet} and Vision Transformer (ViT) variants~\cite{dosovitskiy2020image} trained via the CLIP framework~\cite{radford2021learning}. These models, pretrained on large-scale datasets, capture rich semantic features that are well-suited for downstream classification tasks.

EfficientNet-B3 produces 1536-dimensional embeddings, while ViT models typically output 768 or 512-dimensional vectors. To evaluate trade-offs between representation richness and simulation cost, we experiment with three dimensionality settings: 512, 768, and 1536, across different architectures. For lower-dimensional settings, we apply Principal Component Analysis (PCA) to reduce embedding size while preserving key variance, thereby aligning inputs with the capacity limits of our 16-qubit quantum kernel simulator.

To benchmark performance across embedding strategies, we map each encoder-dimension pair to a shorthand label used throughout our analysis (e.g., \texttt{ViT-B/16-512}, \texttt{EffNet-1536}). Table~\ref{tab:model_mappings} summarizes these configurations, including the native embedding size and dimensionality used in simulation.

As a point of reference, we include two baselines: the \texttt{ Baseline}, based on Chen et al.’s original QSVM using flattened image pixels, and our enhanced version, \texttt{Baseline+}, which incorporates the computational enhancements introduced in Section~\ref{sec:qs_backend}. All remaining models use the same enhanced QSVM backend as \texttt{Baseline+}, differing only in their embedding source and size.

\begin{table}[ht]
\centering
\caption{Model categories and configurations used in the experiments. 
\texttt{Baseline} refers to the original implementation by Chen et al.~\cite{chen2024validating}, using flattened pixel inputs. 
\texttt{Baseline+ (ours)} is our enhanced QSVM implementation using raw pixels. 
All other variants apply the same enhanced QSVM pipeline with different pretrained embeddings.}
\label{tab:model_mappings}
\begin{tabular}{@{}llc@{}}
\toprule
\textbf{Label} & \textbf{Description} & \textbf{Input Dim.} \\
\midrule
\texttt{Baseline}         & Chen et al. with flattened pixels         & 784 \\
\texttt{Baseline+ (ours)}     & Enhanced QSVM pipeline with raw pixels   & 784 \\
\midrule
\texttt{QSVM: EffNet-512}     & EfficientNet-B3 embedding                 & 512 \\
\texttt{QSVM: EffNet-1536}    & EfficientNet-B3 embedding                 & 1536 \\
\texttt{QSVM: ViT-B/32-512}   & ViT-B/32 (CLIP) embedding                 & 512 \\
\texttt{QSVM: ViT-B/16-512}   & ViT-B/16 (CLIP) embedding                 & 512 \\
\texttt{QSVM: ViT-L/14}       & ViT-L/14 (CLIP) embedding                 & 768 \\
\texttt{QSVM: ViT-L/14@336}   & ViT-L/14@336px (CLIP) embedding           & 768 \\
\bottomrule
\end{tabular}
\end{table}
\subsection{Evaluation Methodology}

Model performance is assessed through 5-fold cross-validation to ensure robust statistical evaluation. We measure classification accuracy, precision, F1-score, and Area Under the Curve (AUC) to provide comprehensive performance characterization. Computational efficiency is evaluated by tracking total execution time and peak memory usage during training and evaluation phases.

Classical SVMs are implemented using scikit-learn's SVC with RBF kernel and hyperparameters (C=1.0, $\gamma$='scale', probability=True). All preprocessing steps, including embedding extraction and PCA dimensionality reduction, are identical between classical and quantum approaches. This ensures that performance differences reflect only the kernel computation method rather than data preparation artifacts.

Our evaluation directly contrasts quantum support vector machines against classical SVM baselines using identical feature representations and evaluation protocols. This approach isolates the impact of quantum kernel methods from preprocessing effects, enabling fair assessment of quantum advantage claims. Statistical significance is evaluated through cross-validation consistency and standard deviation analysis.

\subsection{Computational Infrastructure}

All experiments are conducted on NVIDIA A100 Tensor Core GPUs with 80GB HBM2 memory, using CUDA 12.0 and NVIDIA's cuQuantum cuTensorNet backend~\cite{cutensornet} for quantum simulation. This high-performance computing environment ensures consistent benchmarking conditions and enables efficient tensor network contraction for quantum kernel computation. The GPU-accelerated simulation framework allows us to explore larger quantum circuits than would be feasible with CPU-based approaches.

\section{Results and Analysis}

\subsection{Quantum Advantage with Modern Neural Embeddings}

Our central finding demonstrates that quantum support vector machines achieve consistent performance improvements over classical SVMs when using Vision Transformer embeddings, while showing degraded performance with raw pixels or CNN vector embeddings for this specific setting. Table~\ref{tab:qsvm_vs_svc} presents our key results: Quantum models with ViT embeddings achieve accuracy gains up to 4.4\% on MNIST and 8.0\% on Fashion-MNIST, while traditional approaches (raw pixels, EffNet features) show performance degradation. 

\textit{Baseline+} reduced runtime from 4,492 to 3,812 seconds during cross-validation, saving ~680 seconds, and brought peak memory usage down from 44.1GB to 43.5GB. All enhancements to the quantum pipeline, including caching, memory preallocation, and parallel tensor contractions, were built on this enhanced baseline rather than the original, as observed in Table~\ref{tab:results_mnist_fmnist_full}.

\begin{table}[ht!]
\centering
\caption{Quantum vs Classical SVM Performance Comparison. Held-out test accuracy demonstrating quantum advantage with modern neural embeddings. Quantum advantage represents the relative improvement.}
\label{tab:qsvm_vs_svc}
\begin{tabular}{@{}l@{\hskip 4pt}l@{\hskip 6pt}c@{\hskip 6pt}c@{\hskip 6pt}c@{}}
\toprule
\textbf{Dataset} & \textbf{Embedding type} & \textbf{Classic SVM Acc} & \textbf{Baseline+ (ours) Acc} & \textbf{Quantum Advantage} \\
\midrule
\multirow{7}{*}{\scriptsize MNIST}
& Raw Pixels         & \textbf{0.945}  & 0.887  & -6.14  \\
& EffNet-512         & \textbf{0.9694} & 0.935  & -3.55 \\
& EffNet-1536        & \textbf{0.9731} & 0.948  & -2.58 \\
& ViT-B/32-512       & 0.9481 & \textbf{0.990}  & \textbf{+4.42} \\
& ViT-B/16-512       & 0.9544 & \textbf{0.995}  & \textbf{+4.25} \\
& ViT-L/14           & 0.9825 & \textbf{0.990}  & \textbf{+0.76} \\
& ViT-L/14@336-768   & \textbf{0.9838} & \textbf{0.993}  & \textbf{+0.94} \\
\midrule
\multirow{7}{*}{\scriptsize FMNIST}
& Raw Pixels         & \textbf{0.7825} & 0.730  & -6.71 \\
& EffNet-512         & \textbf{0.9172} & 0.887  & -3.29 \\
& EffNet-1536        & \textbf{0.916}  & 0.877  & -4.26 \\
& ViT-B/32-512       & 0.8476 & \textbf{0.900}  & \textbf{+6.18} \\
& ViT-B/16-512       & 0.8332 & \textbf{0.900}  & \textbf{+8.02} \\
& ViT-L/14           & 0.8708 & \textbf{0.897}  & \textbf{+3.01} \\
& ViT-L/14@336-768   & 0.8652 & \textbf{0.900}  & \textbf{+4.02} \\
\bottomrule
\end{tabular}
\end{table}

This quantum advantage emerges specifically with transformer-based representations, revealing a fundamental synergy between quantum kernels and modern neural embeddings. All Vision Transformer variants demonstrate positive quantum advantage, indicating this is not merely an artifact of specific architectural choices but rather reflects a deeper compatibility between quantum feature spaces and transformer-learned representations.

The results highlight the critical importance of feature representation selection in quantum machine learning. Traditional approaches using raw pixels or CNN-based features consistently favor classical methods, while transformer embeddings unlock quantum computational advantages. This finding has important implications for the design of quantum machine learning systems, suggesting that the preprocessing stage is as crucial as the quantum algorithm itself for achieving quantum advantage.

The practical significance of up to 8\% accuracy improvement represents substantial value for real-world applications, particularly in domains where high precision is critical such as medical diagnosis or safety-critical systems. These gains, while seemingly modest, can translate to significant improvements in deployment scenarios where accuracy differences directly impact outcomes.

\begin{figure}[ht!]
    \centering
    \begin{subfigure}[b]{\linewidth}
        \includegraphics[width=\linewidth]{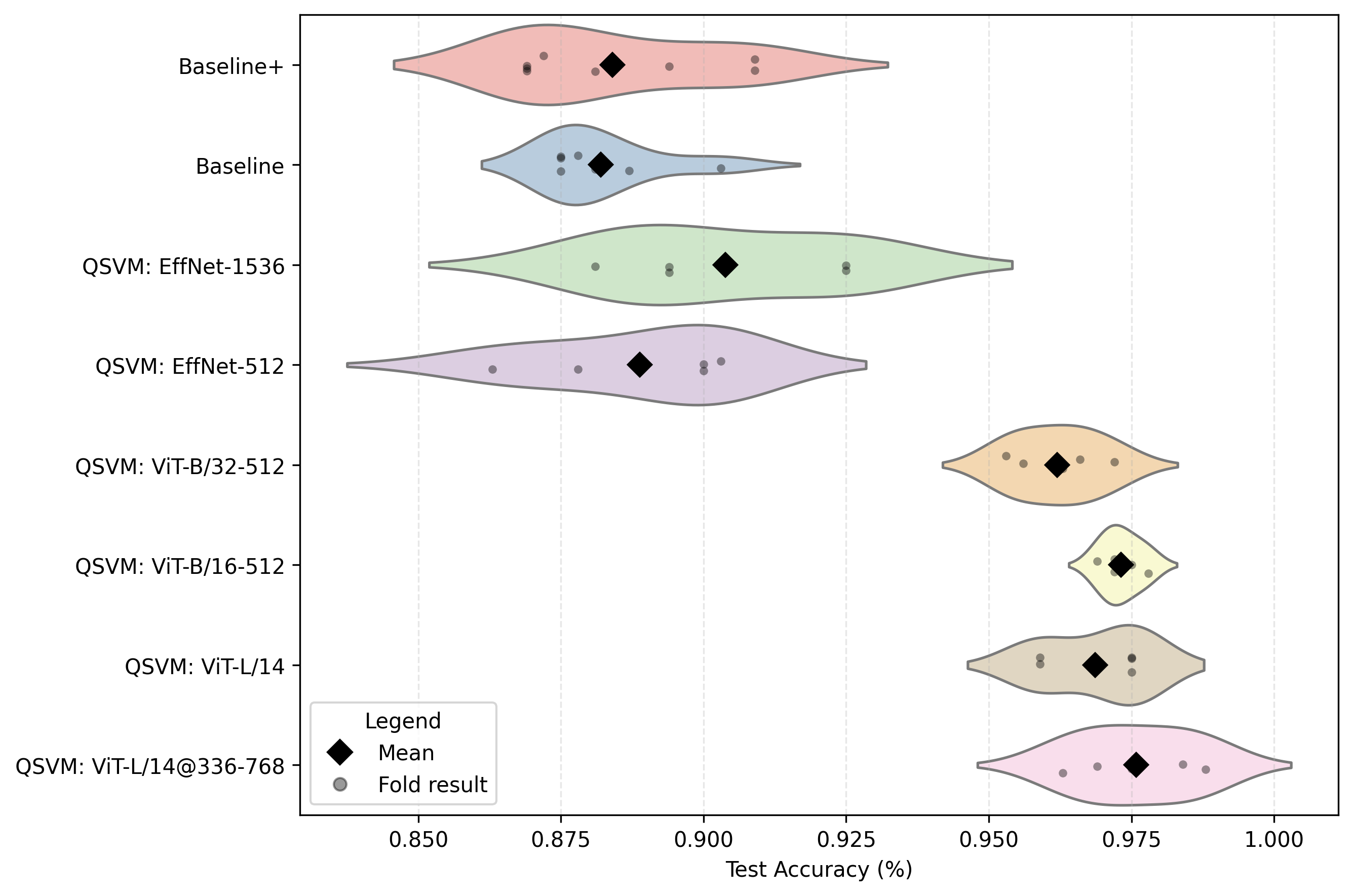}
        \label{fig:mnist32}
    \end{subfigure}
    \caption{Violin plots show the distribution of test accuracy across K-fold cross-validation for MNIST. The width of each violin indicates the density of results; wider sections reflect more frequent accuracy values, helping visualize consistency and variability in model performance.}
    \label{fig:violit_plot_mnist}
\end{figure}

\begin{figure}[ht!]
    \centering
    \begin{subfigure}[b]{\linewidth}
        \includegraphics[width=\linewidth]{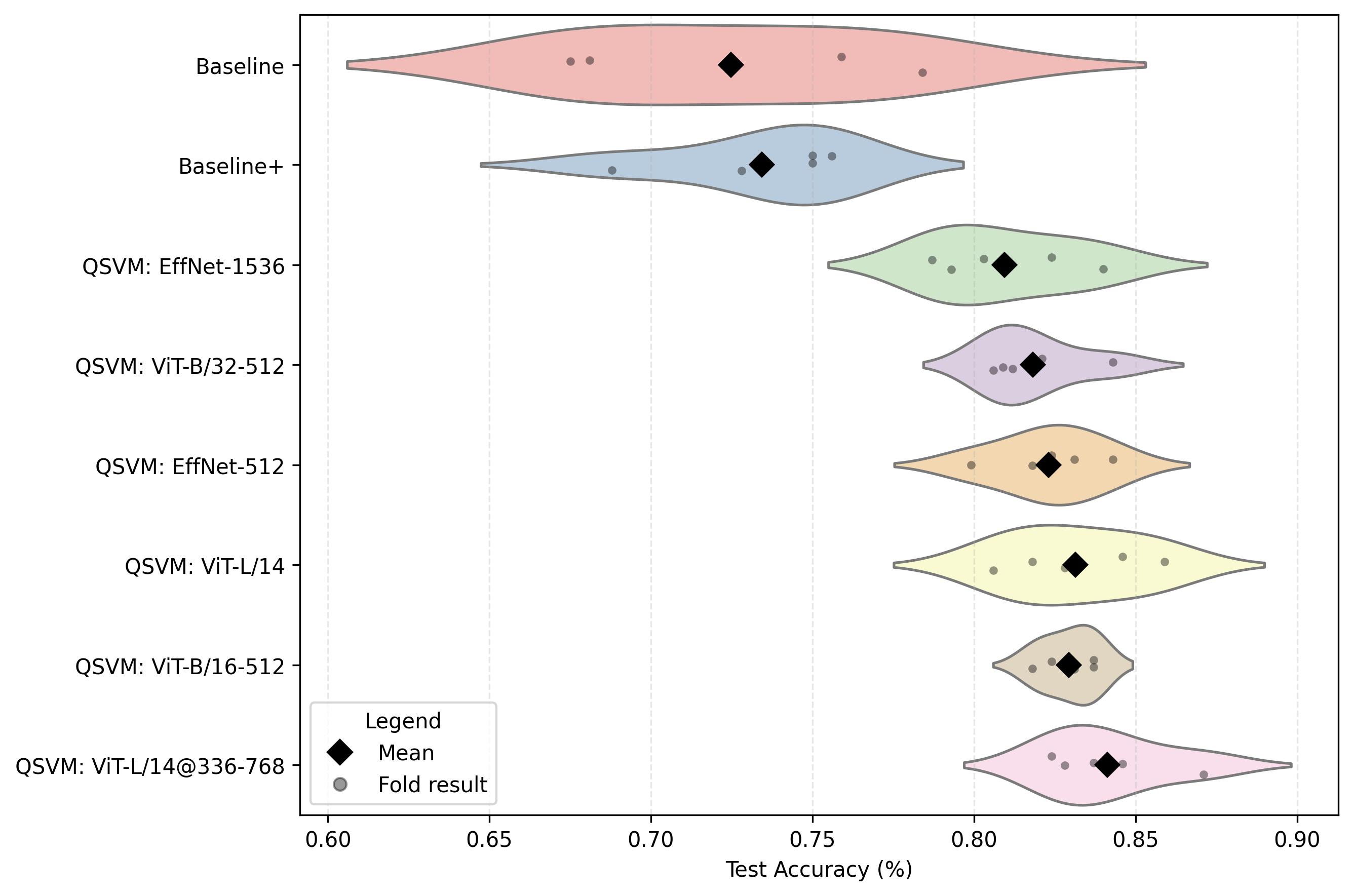}
        \label{fig:mnist32}
    \end{subfigure}
    \caption{Violin plots show the distribution of test accuracy across K-fold cross-validation for FashionMNIST. The width of each violin indicates the density of results; wider sections reflect more frequent accuracy values, helping visualize consistency and variability in model performance.}
    \label{fig:violit_plot_fmnist}
\end{figure}

\begin{figure}[ht!]
    \centering
    \begin{subfigure}[b]{\linewidth}
        \includegraphics[width=\linewidth]{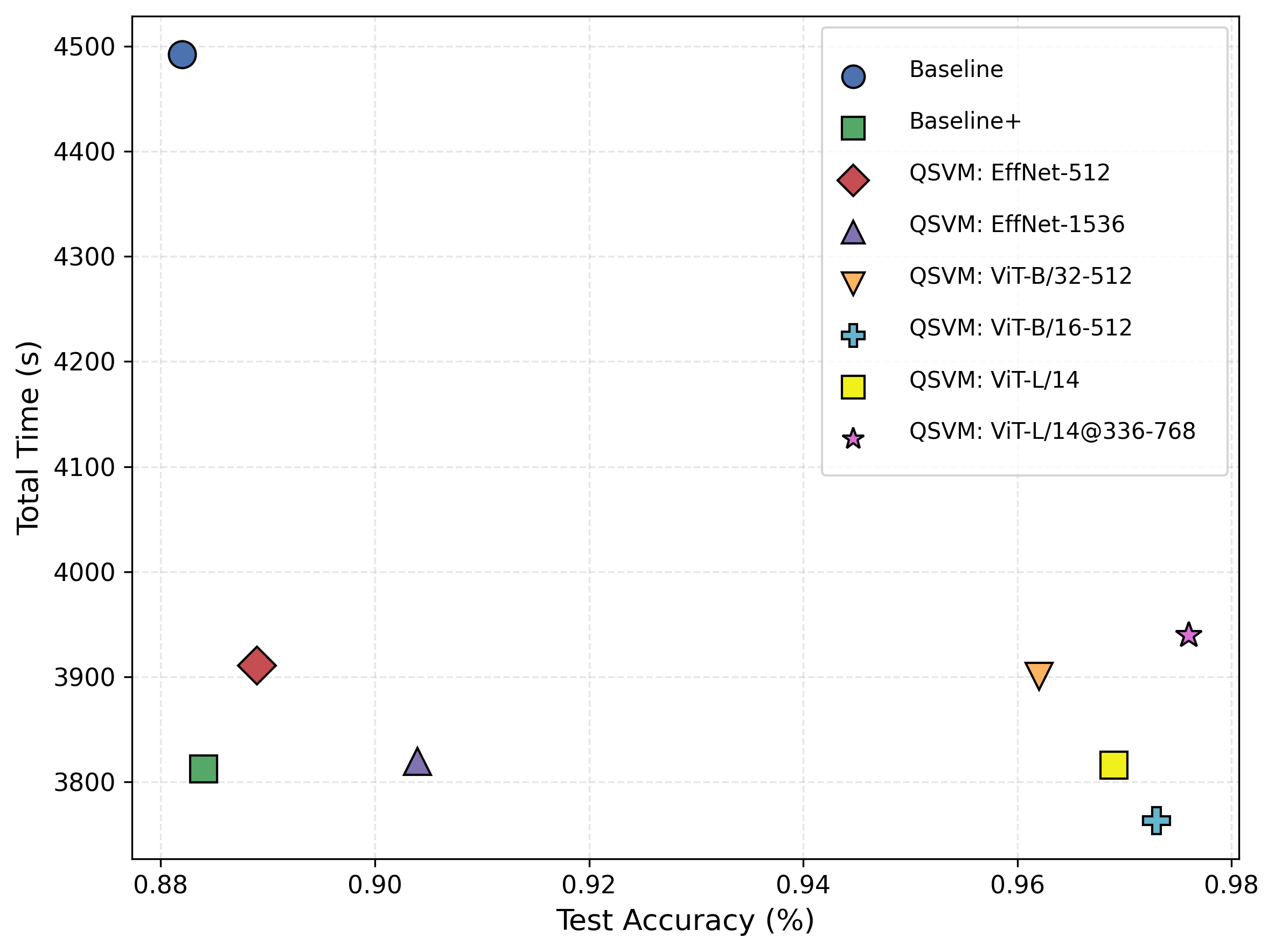}
    \end{subfigure}
    \caption{Comparison of total execution time and test accuracy for different QSVM models for MNIST. The x-axis represents the average test accuracy across K-folds, while the y-axis (log scale) shows the total runtime in seconds. Each point corresponds to a model variant, with horizontal and vertical lines indicating the standard deviation of accuracy and time, respectively.}
    \label{fig:time_vs_accuracy_mnist}
\end{figure}

\begin{figure}[ht!]
    \centering
    \begin{subfigure}[b]{\linewidth}
        \includegraphics[width=\linewidth]{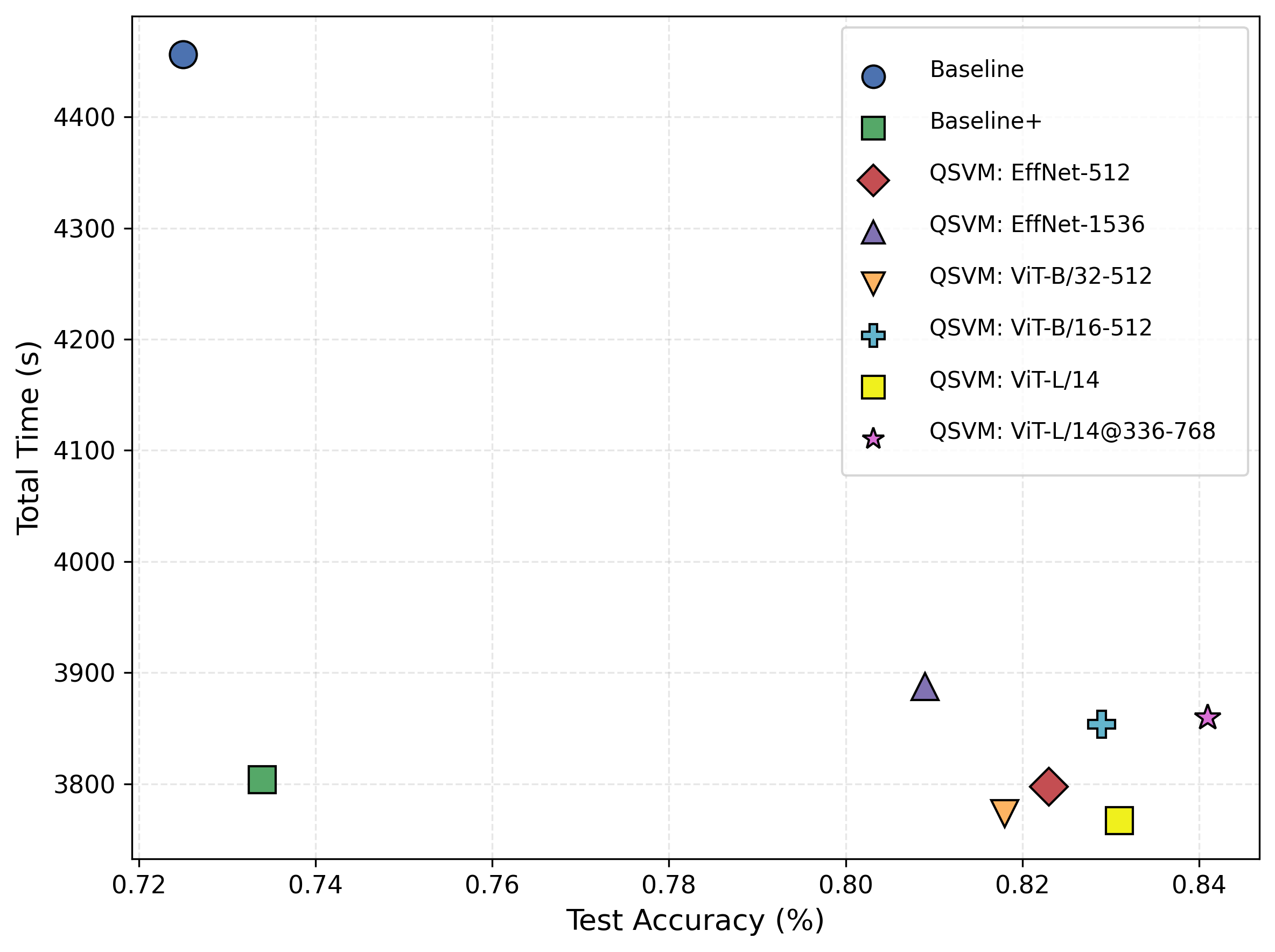}
    \end{subfigure}
    \caption{Comparison of total execution time and test accuracy for different QSVM models for FMNIST. The x-axis represents the average test accuracy across K-folds, while the y-axis (log scale) shows the total runtime in seconds. Each point corresponds to a model variant, with horizontal and vertical lines indicating the standard deviation of accuracy and time, respectively.}
    \label{fig:time_vs_accuracy_fashionmnist}
\end{figure}

\subsection{Cross-Validation Performance Analysis}

Comprehensive 5-fold cross-validation results confirm the robustness of our quantum advantage findings across multiple performance metrics. Table~\ref{tab:results_mnist_fmnist_full} presents detailed performance evaluation including accuracy, precision, F1-score, AUC, runtime, and memory usage for both classical baselines and quantum-enhanced models.

\begin{table}[ht!]
\centering
\caption{Benchmark Results for MNIST and FashionMNIST using 16 qubits. Highest values for accuracy-based metrics and lowest for runtime and memory usage are in \textbf{bold}.}
\label{tab:results_mnist_fmnist_full}
\resizebox{\columnwidth}{!}{%
\begin{tabular}{@{}l@{\hskip 4pt}l@{\hskip 6pt}c@{\hskip 6pt}c@{\hskip 6pt}c@{\hskip 6pt}c@{\hskip 6pt}c@{\hskip 6pt}c@{}}
\toprule
\textbf{Dataset} & \textbf{Model} & \textbf{Test Acc} & \textbf{Precision} & \textbf{F1} & \textbf{AUC} & \textbf{Time (s)} & \textbf{Memory (MB)} \\
\midrule
\multirow{8}{*}{\scriptsize MNIST}
& Baseline & $0.882 \pm 0.010$ & $0.887 \pm 0.010$ & $0.882 \pm 0.011$ & $0.990 \pm 0.004$ & $4492.196 \pm 39.285$ & $44116.842 \pm 25.978$ \\
& Baseline+ & $0.884 \pm 0.018$ & $0.888 \pm 0.019$ & $0.884 \pm 0.018$ & $0.991 \pm 0.004$ & $3812.316 \pm 42.187$ & $43537.845 \pm 22.515$ \\
& QSVM: EffNet-512 & $0.889 \pm 0.018$ & $0.893 \pm 0.015$ & $0.889 \pm 0.017$ & $0.992 \pm 0.003$ & $3910.851 \pm 25.007$ & \textbf{$43506.193 \pm 21.365$} \\
& QSVM: EffNet-1536 & $0.904 \pm 0.020$ & $0.906 \pm 0.019$ & $0.904 \pm 0.020$ & $0.994 \pm 0.003$ & $3819.504 \pm 23.488$ & $43566.972 \pm 22.614$ \\
& QSVM: ViT-B/32-512 & $0.962 \pm 0.008$ & $0.963 \pm 0.007$ & $0.962 \pm 0.008$ & $0.999 \pm 0.000$ & $3900.742 \pm 24.954$ & $43510.314 \pm 21.536$ \\
& QSVM: ViT-B/16-512 & $0.973 \pm 0.003$ & $0.974 \pm 0.003$ & $0.973 \pm 0.003$ & $0.999 \pm 0.000$ & \textbf{$3763.170 \pm 25.646$} & $43513.467 \pm 20.800$ \\
& QSVM: ViT-L/14 & $0.969 \pm 0.009$ & $0.970 \pm 0.008$ & $0.969 \pm 0.008$ & $0.999 \pm 0.001$ & $3816.003 \pm 31.957$ & $43520.979 \pm 18.243$ \\
& \textbf{QSVM: ViT-L/14@336-768} & \textbf{$0.976 \pm 0.010$} & \textbf{$0.977 \pm 0.010$} & \textbf{$0.975 \pm 0.010$} & \textbf{$0.999 \pm 0.001$} & $3939.404 \pm 24.480$ & $43520.375 \pm 22.726$ \\
\midrule
\multirow{8}{*}{\scriptsize FashionMNIST}
& Baseline & $0.725 \pm 0.048$ & $0.723 \pm 0.041$ & $0.716 \pm 0.044$ & $0.963 \pm 0.003$ & $4456.288 \pm 32.991$ & $44086.054 \pm 22.615$ \\
& Baseline+ & $0.734 \pm 0.028$ & $0.727 \pm 0.029$ & $0.723 \pm 0.027$ & $0.963 \pm 0.004$ & $3803.786 \pm 27.142$ & $43510.356 \pm 19.410$ \\
& QSVM: EffNet-512 & $0.823 \pm 0.016$ & $0.823 \pm 0.019$ & $0.818 \pm 0.016$ & $0.980 \pm 0.002$ & $3797.365 \pm 29.575$ & $43256.111 \pm 21.782$ \\
& QSVM: EffNet-1536 & $0.809 \pm 0.022$ & $0.808 \pm 0.020$ & $0.805 \pm 0.020$ & $0.980 \pm 0.004$ & $3887.396 \pm 26.549$ & $43301.836 \pm 17.939$ \\
& QSVM: ViT-B/32-512 & $0.818 \pm 0.015$ & $0.821 \pm 0.014$ & $0.816 \pm 0.015$ & $0.981 \pm 0.002$ & $3773.245 \pm 25.367$ & \textbf{$43250.348 \pm 24.488$} \\
& QSVM: ViT-B/16-512 & $0.829 \pm 0.008$ & $0.831 \pm 0.009$ & $0.827 \pm 0.009$ & $0.982 \pm 0.004$ & $3853.586 \pm 38.180$ & $43258.243 \pm 23.672$ \\
& QSVM: ViT-L/14 & $0.831 \pm 0.021$ & $0.831 \pm 0.022$ & $0.829 \pm 0.022$ & $0.981 \pm 0.003$ & \textbf{$3766.821 \pm 21.742$} & $43266.337 \pm 20.614$ \\
& \textbf{QSVM: ViT-L/14@336-768} & \textbf{$0.841 \pm 0.019$} & \textbf{$0.841 \pm 0.020$} & \textbf{$0.840 \pm 0.020$} & \textbf{$0.983 \pm 0.002$} & $3859.313 \pm 20.656$ & $43265.254 \pm 20.394$ \\
\bottomrule
\end{tabular}%
}
\end{table}

The best-performing quantum model, QSVM using ViT-L/14@336-768, achieves an accuracy of 97.6\% on MNIST and 84.1\% on Fashion-MNIST, clearly surpassing the baselines over pixel information, which level off around 88.2\% and 72.5\%, respectively. The near-perfect AUC scores of 99.9\% across all ViT-based quantum models suggest that these approaches reliably capture discriminative patterns with minimal classification errors.

The consistent improvement across cross-validation folds is especially notable. Accuracy standard deviations remain low, typically between $\pm $0.003 and $\pm $0.020, showing that the observed quantum advantage is stable and reproducible, rather than the result of favorable data splits or initialization. This level of consistency is vital for deployment in practical settings where reliable performance is required.

In addition, the quantum models display strong precision-recall alignment, with precision scores closely tracking overall accuracy across all configurations. This balanced performance suggests that quantum kernels provide meaningful gains in class-level discrimination, rather than boosting accuracy by favoring specific categories.

\begin{figure}[ht!]
    \centering
    \begin{subfigure}[b]{0.4\textwidth}
        \includegraphics[width=\linewidth]{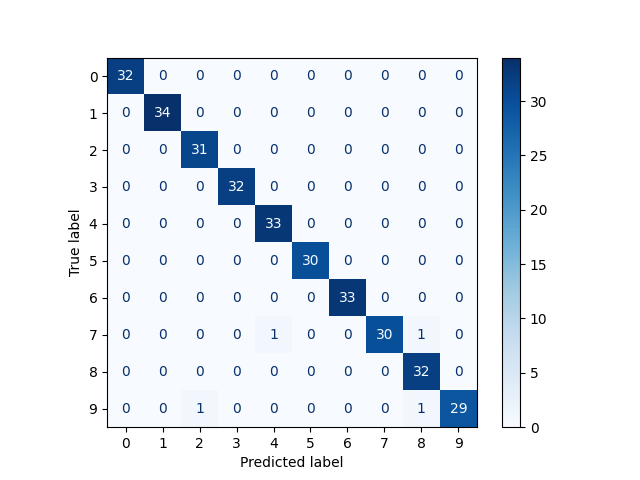}
        \caption{Validation Fold (Best CV Model) for MNIST dataset}
        \label{fig:cv_mnist}
    \end{subfigure}
    \begin{subfigure}[b]{0.4\textwidth}
        \includegraphics[width=\linewidth]{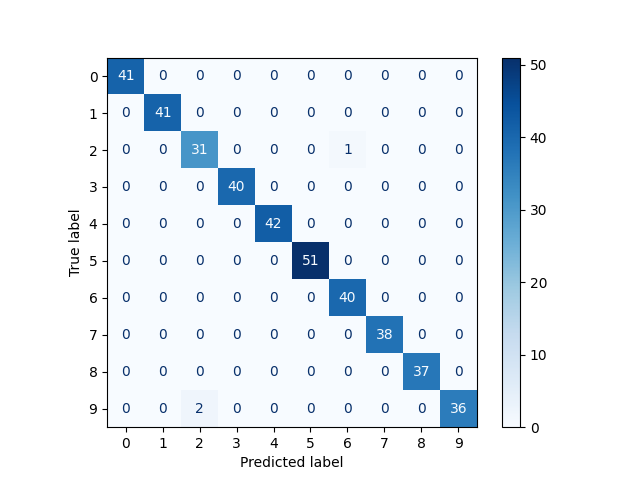}
        \caption{Held-out Test Set (Best CV Model) for MNIST dataset}
        \label{fig:test_mnist}
    \end{subfigure}

    \caption{
        \textbf{(a)} Performance on the validation fold used to select the best model. 
        \textbf{(b)} Performance of that model on the held-out test set. 
        This comparison highlights how well the selected model generalizes to unseen data.
    }
    \label{fig:mnist_cv_test_comparison_mnist}
    
\end{figure}

\begin{figure}[ht!]
    \centering
    \begin{subfigure}[b]{0.4\textwidth}
        \includegraphics[width=\linewidth]{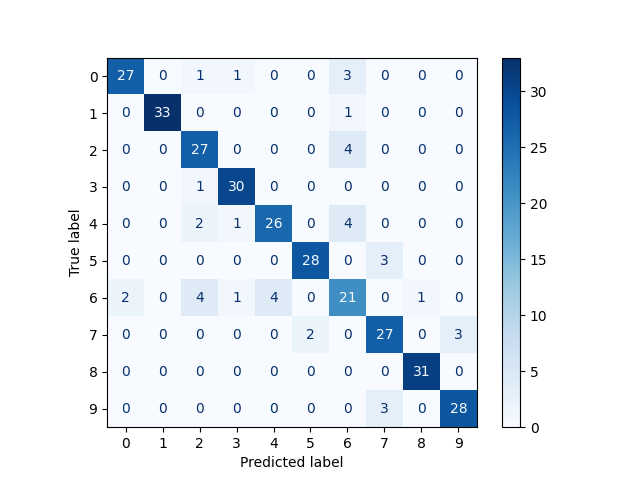}
        \caption{Validation Fold (Best CV Model) for Fashion MNIST dataset}
        \label{fig:cv_mnist}
    \end{subfigure}
    \begin{subfigure}[b]{0.4\textwidth}
        \includegraphics[width=\linewidth]{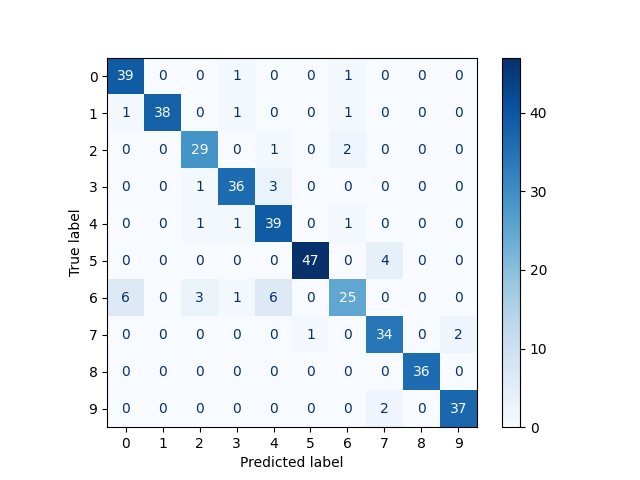}
        \caption{Held-out Test Set (Best CV Model) for Fashion MNIST dataset}
        \label{fig:test_mnist}
    \end{subfigure}

    \caption{
        \textbf{(a)} Performance on the validation fold used to select the best model. 
        \textbf{(b)} Performance of that model on the held-out test set. 
        This comparison highlights how well the selected model generalizes to unseen data.
    }
    \label{fig:mnist_cv_test_comparison_fmnist}
    
\end{figure}

The confusion matrices in Figures~\ref{fig:mnist_cv_test_comparison_mnist} and \ref{fig:mnist_cv_test_comparison_fmnist} further illustrate the generalization power of our top model, QSVM with ViT-L/14@336-768, showing alignment between cross-validation and held-out test results. Generalization is stronger for MNIST than for Fashion-MNIST. The clear diagonal structure and few off-diagonal errors confirm that the high accuracy reflects true performance across all digit classes, not just select ones.

Violin plots in Figures~\ref{fig:violit_plot_mnist} and~\ref{fig:violit_plot_fmnist} visualize test accuracy distributions over cross-validation folds. QSVM models with ViT embeddings, including ViT-B/16, ViT-L/14, and ViT-L/14@336, consistently achieve higher average accuracies and lower variance compared to both baselines and EfficientNet-based QSVMs. ViT-B/16-512 and ViT-L/14@336-768 show especially narrow, high-accuracy distributions on MNIST, while baseline and EfficientNet models display wider, more variable spreads, particularly on Fashion-MNIST. These results highlight the advantage of transformer embeddings in delivering stable and accurate quantum classification across tasks of varying complexity.

\subsection{Computational Efficiency and Scalability Analysis}

Our embedding-enhanced quantum models demonstrate strong accuracy while maintaining reasonable computational demands for quantum simulations. Most ViT-based quantum configurations complete training and evaluation in approximately 3,800 seconds with consistent memory usage around 43GB, as detailed in Table~\ref{tab:results_mnist_fmnist_full}. While these runtimes may appear substantial, they represent a significant improvement over prior quantum simulations and are reasonable given the high-dimensional embedding spaces and tensor contraction overhead inherent to quantum kernel methods.

Among the top-performing models, QSVM with ViT-B/16-512 offers the optimal balance between performance and efficiency, achieving 97.3\% accuracy with the fastest runtime of 3,763 seconds. Figures~\ref{fig:time_vs_accuracy_mnist} and~\ref{fig:time_vs_accuracy_fashionmnist} illustrate the trade-offs between computational cost and classification performance. Vision Transformer-based models consistently achieve top-tier accuracy with moderate computational requirements, while EfficientNet configurations provide competitive accuracy with reduced resource demands. The original \textit{Baseline} model shows the least favorable performance-efficiency balance, while \textit{Baseline+} demonstrates consistent runtime improvements.

These results confirm that embedding-enhanced quantum models offer practical scalability alongside significant accuracy gains. Vision Transformer embeddings clearly outperform EfficientNet-B3 across both datasets, and the computational overhead of quantum kernel methods is effectively mitigated by the reduced dataset sizes achieved through strategic data distillation, making these approaches viable for real-world deployment.

\section{Discussion}

This study demonstrates that quantum advantage in machine learning emerges not from quantum algorithms alone, but from the strategic synergy between quantum kernels and appropriate feature representations. Our central finding reveals that Vision Transformer embeddings uniquely unlock quantum advantage, achieving up to 8.02\% accuracy improvements over classical SVMs, while CNN features and raw pixels consistently favor classical approaches.

Although quantum simulations demand substantial computational resources (approximately 3,800 seconds for training), this investment proves justified in high-precision applications where accuracy improvements directly translate to enhanced outcomes, particularly in medical diagnosis, safety-critical systems, and fraud detection scenarios. The computational overhead becomes advantageous compared to scaling classical approaches for similar accuracy gains, which typically require exponentially larger datasets or increasingly complex architectures. Our strategic data distillation effectively reduces problem complexity from $\mathcal{O}(70000^2)$ to $\mathcal{O}(1600^2)$ kernel evaluations, making quantum kernel methods tractable while preserving essential dataset characteristics.

Our framework demonstrates that quantum machine learning achieves scalability through intelligent preprocessing, with distillation parameters easily customizable based on available computational resources. This adaptability enables flexible deployment across diverse scenarios, from resource-constrained environments utilizing smaller distilled datasets to high-performance settings leveraging full quantum simulation capabilities. The embedding-aware approach establishes a practical pathway toward quantum advantage that becomes increasingly favorable as quantum hardware continues to mature.

\section{Conclusion}

We present an embedding-aware quantum-classical framework that systematically addresses scalability challenges in quantum machine learning by strategically combining class-balanced data distillation with pretrained embeddings. Building upon Chen et al.'s GPU-accelerated quantum kernel method~\cite{chen2024validating}, our pipeline successfully reduces computational complexity while achieving measurable performance improvements over classical baselines.

Our work delivers the first systematic evidence that quantum kernel advantage depends critically on embedding choice, revealing fundamental compatibility between transformer attention mechanisms and quantum feature spaces. Through 16-qubit tensor network simulation, we demonstrate consistent quantum advantages using ViT embeddings across MNIST (up to 4.42\% improvement) and Fashion-MNIST (up to 8.02\% improvement), while observing performance degradation with CNN-based features.

The embedding-aware QSVM framework enables practical quantum machine learning deployment through configurable data distillation and hardware-adaptive preprocessing. Compared to raw inputs, structured transformer embeddings consistently deliver superior accuracy and generalization, effectively supporting real-world applications in high-dimensional classification tasks where precision remains critical.

\textbf{Limitations and Future Directions.} Several limitations require attention for broader impact. Our evaluation concentrates on relatively simple visual classification benchmarks (MNIST, Fashion-MNIST); validation on complex datasets such as CIFAR-10, medical imaging, or domain-specific applications remains necessary to assess generalization. The theoretical foundations explaining transformer-quantum synergy remain largely underexplored, presenting compelling opportunities for fundamental research.

Future work should pursue automated embedding and kernel selection strategies~\cite{Incudini_2024} to eliminate manual hyperparameter tuning, explore sophisticated dimensionality reduction techniques beyond PCA to better preserve semantic information, and develop optimized quantum circuit designs~\cite{sünkel2023ga4qcogeneticalgorithmquantum} for enhanced computational efficiency. Expanding empirical validation to medical imaging and other high-dimensional domains will prove critical for demonstrating broader practical utility.

This work establishes that achieving quantum advantage requires careful algorithm-representation co-design rather than naive application of quantum methods. Our embedding-aware framework provides both immediate practical value for precision-critical applications and a scalable foundation for quantum machine learning that effectively leverages modern neural architectures. As quantum hardware continues to mature, this approach offers a viable pathway toward practical quantum advantage in real-world machine learning applications.

\section{Data and Code Availability}

The code used in this study is publicly available at: \href{https://github.com/sebasmos/QuantumVE}{https://github.com/sebasmos/QuantumVE}.

\section{Acknowledgments}

This work was supported by the Google Cloud Research Credits program under the award number GCP19980904.

  

\bibliography{paper}

\end{document}